\documentclass[preprint,prd,noshowpacs]{revtex4}
\usepackage{amssymb}
\usepackage{amsmath} 

\begin{document} 

\title{De Sitter space and perpetuum mobile} 

\author{Emil T. Akhmedov}
\email{akhmedov@itep.ru}
\affiliation{ITEP, 117218 Russia, Moscow,
B. Cheremushkinskaya str. 25; \\
Moscow Institute of Physics and Technology, Dolgoprudny, Russia}
\author{P. V. Buividovich}
\email{buividovich@tut.by} \affiliation{JIPNR, National Academy of
Science, 220109 Belarus, Minsk, Acad. Krasin str. 99;
\\ ITEP, 117218 Russia, Moscow, B. Cheremushkinskaya str. 25}
\author{Douglas A. Singleton}
\email{dougs@csufresno.edu}
\affiliation{Physics Department, CSU Fresno, Fresno, CA 93740 USA; \\
Institute of Gravitation and Cosmology, Peoples' Friendship
University of Russia, Moscow 117198, Russia} 

\date{\today} 

\begin{abstract}
We give general arguments that any interacting non--conformal {\it
classical} field theory in de Sitter space leads to the
possibility of constructing a perpetuum mobile. The arguments are
based on the observation that massive free falling particles can
radiate other massive particles on the classical level as seen by
the free falling observer. The intensity of the radiation process
is non-zero even for particles with any finite mass, i.e. with a
wavelength which is within the causal domain. Hence, we conclude that
either de Sitter space can
not exist eternally or that one can build a perpetuum mobile. \\
\end{abstract} 

\maketitle 

\section{Introduction} 

 It is generally believed that de Sitter space is stable at least
on the classical level. The belief relies on the following
arguments: (i) de Sitter space has a big isometry group
\cite{BornerDurr, Kim:2002uz}; (ii) there are {\it
no} exponentially growing linearized fluctuations over de Sitter
space \cite{Lifshitz:1945du}. 

The question is whether these arguments are sufficient to prove
the classical stability of de Sitter space. In this note we argue
that the answer on this question is negative. Indeed the situation
changes if one turns on interactions. Let us ask the following
question of a classical \footnote{We stress that the reason why we call
our considerations classical is because in this paper we are just dealing
with solutions of non--linear, classical wave equations.} interacting field 
theory on de
Sitter background: ``Does an inertially moving charged particle in de
Sitter space emit radiation or not?''. Because the space in
question is conformaly flat, we consider a field theory which is
not conformaly invariant, otherwise the behavior of fields is not
much different from fields in Minkowski space. 

Calculating the {\it classical} amplitude \footnote{We explain in the
main body of the text exactly what we mean by ``classical radiation''
and ``classical amplitude''.} of the corresponding
process and observing that it is not zero, we obtain an
affirmative answer to the above question. Our point is that even
in a classical field theory on de Sitter space, massive particles
can radiate other massive particles with a wavelength within the
causal domain. 

{\it Quantum} particle production by another particle in de Sitter
space was addressed in \cite{Myrhvold, Akhmedov:2008pu,
Tsamis:1996qq, Bros:2009bz}. Classical electromagnetic radiation
was considered in \cite{Tsaregorodtsev:1998mq},\cite{Poisson}.
On general physical grounds one can expect that the characteristic
wavelength of the radiation of the massless fields and the wave creation 
length in such a
process are of the order of the size of the cosmological horizon.
This is one of the reasons why we consider radiation of massive
particles with wavelengths smaller than the cosmological horizon.
Our main point in this note is to show that there is a problem with
the stability of de Sitter space even on the classical level -- if
the vacuum energy is truly fixed then one can create a perpetuum mobile. 

\section{Field theory in de Sitter space} 

The $D$--dimensional de Sitter space is a hyperboloid
\cite{BornerDurr, Bousso:2001mw, Kim:2002uz}
\begin{equation}
\label{hyperb}
 - z_0^2 + \sum_{i=1}^D z_i^2 = 1
\end{equation}
in $(D+1)$--dimensional Minkowski space. In this note we set the
radius of de Sitter space to 1 and consider $D\ge 3$. One can see
explicitly that the isometry of de Sitter space is the
$SO(D,1)$ symmetry group of the background $(D+1)$--dimensional space.
The presence of this large isometry is one of the arguments
favoring the stability of de Sitter space --- a space with a
running vacuum energy would be less symmetric. The reason why such an
argument is not sufficient is that in field theory it frequently happens that
a less symmetric state is energetically more favorable than the more
symmetric one. The classic example is the spontaneous symmetry breaking. 

The specific induced metric on the hyperboloid, used below, is:
\begin{equation}
ds^2 = - dt^2 + \cosh^2 t \, d\Omega^2_{D-1},
\label{global}
\end{equation}
where $d\Omega^2_{D-1}$ is the metric on the unit
$(D-1)$--dimensional sphere. These ``global'' coordinates cover
the entire de Sitter space, and are those seen by inertial
observers, since the coordinate time in this metric coincides with
the proper time. All the results in this paper are those seen by an inertial
observer. 

Consider linearized fluctuations in de Sitter space. For
simplicity and because we will use it below, we take a scalar field,
$\Phi$, with arbitrary mass,
$m$. The Klein--Gordon equation in the global coordinates is:
\begin{equation}
\left(\partial_t^2 + (D-2) \tanh t \, \partial_t + m^2 -
\frac{\Delta_{D-1}(\Omega)}{\cosh^2 t}\right) \, \Phi = 0,
\label{waveglobal}
\end{equation}
$\Delta_{D-1}(\Omega)$ is the Laplacian on the
$(D-1)$--dimensional sphere. 

Just by looking at the asymptotic form of \eqref{waveglobal} as
$t\to\pm\infty$, one can see that none of its solutions grows
exponentially with time. Linearized fluctuations of the metric
obey equations which are very similar to \eqref{waveglobal} with
the zero mass. Hence, similar arguments lead to the conclusion that
they do not grow exponentially with time and
are not able to change the background de Sitter metric
\cite{Lifshitz:1945du}. These arguments are valid on the linearized
level, while we would like to consider an interacting theory. 

To address our question we have to define what is meant
by a particle in de Sitter space. Equation \eqref{waveglobal} is
linear thus it seems that one could consider as particles any basis of
solutions, because they obey the superposition principle. However,
this is not quite correct. For example by taking an arbitrary
basis of solutions of the Klein--Gordon equation (e.g. linear
combination of positive and negative energy harmonics) in
Minkowski space one can obtain radiation on mass--shell, which is
known to be wrong. One must define particles as positive energy
excitations over the Poincare invariant vacuum state to ensure
that there is no radiation on mass--shell in Minkowski space (i.e.
no radiation from a particle moving with constant velocity). We clarify this
observation below. 

It is worth stressing here that one can use {\it any} basis of solutions of
\eqref{waveglobal}, but the resulting answer to the question posed in the
introduction will be the same. 

 Using the separation of variables $\Phi_{j\, n}(t, \Omega) =
\varphi_j(t) \, Y_{j \, n}(\Omega)$ one can find the basis of solutions of
\eqref{waveglobal} \cite{BornerDurr, Mottola:1984ar}. Here
$\Delta_{D-1}(\Omega) \, Y_{j\, n}(\Omega) = - j(j + D-2)\, Y_{j\, n}(\Omega)$, and $n$
is the multi--index $(n_1,\dots, n_{D-2})$. 

The properties of these spherical harmonics $Y_{j\, n}(\Omega)$
are given in \cite{Bousso:2001mw}. The field $\varphi_j(t)$ obeys
an equation following from \eqref{waveglobal}. This equation has
two distinguished complete sets of solutions: the in-- and
out--modes \cite{Bousso:2001mw, Mottola:1984ar}. The complete set
of in--modes is
\begin{equation}
\varphi_j^{\pm}(t) \propto \cosh^j(t) \,
e^{\left(j + \frac{D-1}{2} \mp i \, \mu\right)\, t} \, F\left(j +
\frac{D-1}{2}, \, j + \frac{D-1}{2} \mp i\,\mu; 1 \mp i\, \mu; -
e^{2\,t} \right)
\label{solution}
\end{equation}
where $ \mu = \sqrt{m^2
 -\left(\frac{D-1}{2}\right)^2}$ and $F(a,b;c;z)$ is the
hypergeometric function. The solution \eqref{solution} can be
continued to the case when $m < (D-1)/2$. 

The reason for the name of these modes is that they behave at past
infinity ($t\to-\infty$) as $\varphi_j^{\pm} \rightarrow
e^{\left(\frac{D-1}{2} \mp i \, \mu\right)\, t}$ and, hence,
diagonalize the classical free Hamiltonian of the scalar field theory in
this region of space--time. At future infinity ($t\to +\infty$)
they behave as $\varphi_j^{\pm} \rightarrow
e^{-\frac{D-1}{2}\,t}\left(c_1 \, e^{\mp i \,\mu \, t} + c_2 \,
e^{\pm i \,\mu \, t}\right)$ with some non--zero (for even
\footnote{If $D$ is odd, however, $c_2 = 0$ and it is believed that there is
no
ambiguity in the choice of the de Sitter invariant vacuum state
\cite{Bousso:2001mw}.} $D$) complex constants $c_1$ and $c_2$.
Hence, for even $D$ at future infinity they do {\it not}
diagonalize the free Hamiltonian. At future infinity the free
Hamiltonian is diagonalized by the out--modes
$\bar{\varphi}^{\pm}_j(t)$, which are related to the in--modes by
$\bar{\varphi}^{\pm}_j(t) = \left(\varphi^{\pm}_j(- t)\right)^*.$ 

The vitally important fact for our considerations below is that, unlike
Minkowski and AdS spaces, there is {\it no} set of solutions of
\eqref{waveglobal}, which diagonalizes the classical free Hamiltonian
in de Sitter space for all times.
This is true in any dimension although for odd $D$ the same set of harmonics
does diagonalizes the free Hamiltonian both at past and future infinities.

\section{What is meant by classical radiation} 

Although our arguments
are general, for concreteness we consider equations of
motion in a Yukawa type theory describing the interaction of a massive
($M$) fermion, $\Psi$, with a massive ($m$) boson, $\Phi$: 

\begin{eqnarray}\label{system}
\left[\Delta(g) + m^2\right]\,\Phi = \lambda \, \bar{\Psi}\,\Psi \nonumber
\\
\left[\hat{D} + M\right]\,\Psi = \lambda\,\Phi \bar{\Psi}.
\end{eqnarray}
As well there is the complex conjugate equation for the fermion field.
Here $\Delta(g)$ is the d'Alembertian for the de Sitter metric,
$\hat{D}$ is the Dirac operator for the same metric, and $\lambda$ is
the interaction constant. The metric is taken to be non--dynamical.
We have chosen a Yukawa type theory in order to avoid
the possible counterarguments against $\phi^3$ theory, which are based on
the absence of a minimum for the potential energy for this theory. 

First let us consider this field theory in Minkowski space. The
simplest solution of the above system
of equations is $\Psi=0$ and $\Phi=0$ (and a flat metric solution of the
Einstein--Hilbert equations), which corresponds to empty space.
If $\lambda$ were zero one could excite the $\Phi$ and $\Psi$ fields to
see that there is no exponentially growing mode in Minkowski or in
de Sitter space. 

We are interested here in what happens when one adiabatically turns on 
$\lambda$ at past infinity and switches it of at the future infinity.
Suppose we initial set $\Phi$ to zero, but excite the field $\Psi$
i.e. we add one mass--shell harmonic of this field on top of the
de Sitter background. Then according to the system of equations
\eqref{system} the solution for $\Phi$ to {\it leading order} in $\lambda$
is as follows 

\begin{eqnarray}\label{solution11}
\Phi(x) = \lambda\, \int d^Dy G_{R}(x,y)\, \bar{\Psi}(y)\,\Psi(y).
\end{eqnarray}
Here $G_R(x,y)$ is the retarded Green function for the massive scalar field.
Its explicit form is not necessary for our further considerations. We just
need to know that it can always be represented as 

\begin{eqnarray}\label{propag}
G_R(x,y) = \theta(x_0-y_0)\,\sum_\kappa \psi_\kappa(x)\,\psi_\kappa^*(y),
\end{eqnarray}
where $\psi_\kappa(x)$ is the basis of the solutions (i.e. mass--shell 
harmonics)
of the Klein--Gordon equation with proper normalization. 

Now the mass--shell harmonics -- i.e. solutions of \eqref{system} --
in flat space for both bosons and fermions
are proportional to the plane wave $e^{i\,\vec{p}\,\vec{x}}$ and their time
dependence is given by $e^{-i\,\sqrt{\vec{p}^2 + {\rm mass}^2}\,t}$ i.e. in 
Minkowski space the
role of $\kappa$ is played by the momentum, $\vec{p}$. 

Thus, our solution for the scalar field is given by 

\begin{eqnarray}\label{solution1}
\Phi(x) \propto \lambda\,\int d^{D-1}q 
\frac{e^{i\,q_\mu\,x^\mu}}{\sqrt{q^2+m^2}}
\, \int_{-\infty}^{x_0} dy_0\int d^{D-1}y  \, e^{i(p - q - k)_\mu\,y^\mu}.
\end{eqnarray}
Where we have explicitly substituted the Fourier expansion of the Green
function (the plane waves for the mass--shell harmonics) and ignored the
spinor pre--factors for the Fermi fields. 

Now consider the last integral in this formula when $x_0\to+\infty$.
It is just the classical amplitude for the
radiation for this theory in {\it Minkowski} space. We will explain why this 
is so in a moment. It is proportional to: 

\begin{equation}
A \propto \int d^Dy \, e^{i(p - q - k)y} \propto
\delta^{(D)} \left(p - q - k\right), \label{Mink}
\end{equation}
In this formula $p$ and $k$ are the momenta
of the fermion before and after the interaction process and $q$ is the
momentum of the emitted boson field. The $\delta$--function just
imposes energy-momentum conservation at the vertex --- $p=k+q$. Note
that in the classical theory the solution for $\bar{\Psi}$ is just the complex 
conjugate of $\Psi$ and, hence, $p=k$, but we keep them more general. 

All of the three momenta in the amplitude are
on--shell, i.e. $k^2-M^2=p^2 - M^2 = 0$ and $q^2 - m^2 = 0$. Due to the
latter relations the argument of the $\delta$-function is never zero. Hence, 
the amplitude is zero for all allowed $p,k$ and $q$. That means that if,
in the past infinity, we excite one positive energy mass--shell harmonic of 
the fermion field it will not excite (emit) the boson field at future infinity. We 
will have just the single
fermion plane wave excitation over Minkowski space. Note that it {\it does 
not} mean that on the Minkowski
background the system \eqref{system}
can have single mass--shell wave solution for $\Psi$ with $\Phi$ set to 
zero: at intermediate
times the $\Phi$ field is not zero but it does vanish as $x_0\to+\infty$ 
(where in fact $\lambda$ is adiabatically switched off). 

One can show that the same thing happens at higher orders in 
$\lambda$. Thus, as expected, that there is no
classical radiation from a free floating particle in Minkowski
space. Coming back to the discussion of the proper choice of
vacuum and positive energy harmonics, let us stress that if one
had chosen as the positive energy harmonics a linear combination
of $e^{-i\,p\,x}$ and $e^{i\,p\,x}$ with various $p$'s instead of
a single $e^{-i\,p\,x}$, then one would have obtained a non--zero
amplitude instead of \eqref{Mink}, which is known to be physically
incorrect. That is why we are careful with the choice of vacuum
and positive harmonics. 

Note that performing similar calculations in the same theory, but
in the Einstein static space, $ds^2 = - dt^2 + R^2 \,
d\Omega_{D-1}^2$, $R={\rm const.}$, one observes
\cite{Akhmedov:2008pu} that an amplitude such as that in
\eqref{Mink} is zero, similar to the Minkowski space case. 

\section{Radiation in de Sitter space} 

Now we will show that the situation in de Sitter space is quite different.
Again we would like to excite the theory with a single fermion
mass--shell harmonic and see whether this does or does not lead to the 
excitation of the boson field. 

We choose as the mass--shell harmonics the above mentioned in--solutions.
However it is important to stress that whatever harmonics we choose
the classical amplitude will always be non--zero and, hence, the excitation
of the boson field will always occur. The fact that the amplitude
given below does not vanish
is closely related to the fact that one can not find a basis of harmonics in
de Sitter space which diagonalizes the classical free Hamiltonian for all  times. 

Taking the same route as above from the equation \eqref{system} to
\eqref{Mink} we obtain instead of \eqref{Mink} the following classical 
amplitude: 

\begin{eqnarray}
A \propto \int d\Omega Y_{j_1\,n_1}(\Omega)\,
Y^*_{j_2\,n_2}(\Omega)\, Y^*_{j_3\,n_3}(\Omega)\,
\int_{-\infty}^{+\infty} dt \cosh^{D-1}(t) \times \nonumber \\
\times \left[\cosh^{j_1}(t) \, e^{\left(j_1 + \frac{D-1}{2} + i \,
\mu_1\right)\, t} F\left(j_1 + \frac{D-1}{2}, j_1 + \frac{D-1}{2}
+ i \, \mu_1; 1 + i\, \mu_1; -e^{2\,t} \right)\right]\times \nonumber \\
\times \left[\cosh^{j_2}(t) \, e^{\left(j_2 +
\frac{D-1}{2} - i \, \mu_1\right)\, t} F\left(j_2 + \frac{D-1}{2},
j_2 + \frac{D-1}{2} - i \, \mu_1; 1 - i\, \mu_1; -e^{2\,t}
\right)\right]\times \nonumber \\
\times \left[\cosh^{j_3}(t) \,
e^{\left(j_3 + \frac{D-1}{2} - i \, \mu_2\right)\, t} F\left(j_3 +
\frac{D-1}{2}, j_3 + \frac{D-1}{2} - i \, \mu_2; 1 - i\, \mu_2;
 -e^{2\,t} \right)\right]
\label{ampmasshel}
\end{eqnarray}
where $\mu_1 = \sqrt{M^2 - \left(\frac{D-1}{2}\right)^2}$, $\mu_2
= \sqrt{m^2 - \left(\frac{D-1}{2}\right)^2}$. Note that
this amplitude is invariant under general
covariance by construction. 

If either mass, $m$ or $M$, is vanishing, the time integral in the
amplitude (\ref{ampmasshel}) is divergent \cite{Akhmedov:2008pu} (we shortly
discuss the case of $m=0$ in the concluding section).
However, if we keep both masses $M$ and $m$ non--zero, this time
integral is convergent \cite{Akhmedov:2008pu}, because the
harmonics under the integral exponentially decay as
$t\to\pm\infty$. 

The simplest way to see that the amplitude \eqref{ampmasshel} is
not zero is to note that it is the analytical continuation (from
the $D$--dimensional sphere to $D$--dimensional de Sitter space)
of the generalized $3j$ symbol on the $D$--dimensional sphere.
Explicit numerical calculation of the time integral in
\eqref{ampmasshel} shows that it is not zero for $|j_2 - j_3| \leq
j_1 \leq j_2 + j_3$ \cite{Akhmedov:2008pu}, i.e. where the $3j$
symbol in \eqref{ampmasshel} is non--zero. The amplitude
\eqref{ampmasshel} is non--zero for any large, but finite mass.
Note that the amplitude is non--zero even if $j_1=j_2$,
i.e. when $\bar{\Psi}$ is taken as just the complex conjugate of $\Psi$. 

Hence, in \eqref{solution} we have excited the field $\Phi$ at future 
infinity due to the presence of the mass-shell field $\Psi$ at the initial stage.
Such a process is naturally identified as radiation ---
such a definition works universally for all types of space--times.
If we take into account higher corrections in $\lambda$ then we
observe the creation of many harmonics of the field $\Phi$. 

The intensity of the formation of the concrete mode of the field
$\Phi$ is proportional to the square modulus of the above defined
classical amplitude. The process is indeed classical, because we are just
studying the solution of the non--linear, classical equations of motion
and do not take into account any quantum field theoretic effects. 

\section{Conclusions} 

We conclude that in a classical field theory in de Sitter
space, massive particles can radiate other massive particles. The
only force which acts on the particles is due to the gravitational
background induced by the vacuum (``dark'') energy.  It is the {\it
inertially} moving observer which sees the radiation from the free floating 
bodies in de Sitter
space, because the charged body and inertial observer accelerate
away from each other along geodesics in de Sitter space. The
important observation for our discussion is that the only force
which acts both on the charged body and the inertial observer is the
gravitational one, which is due to the vacuum energy. 

One could also calculate the classical energy momentum flux
of various {\it massless fields} from the corresponding free falling
charges in de Sitter space \cite{Sadofyev}. It is straightforward to see
that there is no electromagnetic radiation, because Maxwell's theory is
conformal and its classical physics in de Sitter space is not different from 
that in Minkowski space. Rather unexpectedly for us there is no gravitational 
radiation \cite{Sadofyev}. However, there is the radiation of massless minimally 
coupled field, $\phi$, which confirms the observations of \cite{Poisson}. But we 
disagree with the conclusions of the latter paper that the radiation will stop, when the 
effective mass, $m - q\,\phi$ (where $q$ is the scalar charge) vanishes \cite{Sadofyev}. It 
is declared in \cite{Poisson} that the charged particle just disappears when the effective 
mass vanishes. We can find no reason why the effective mass can not be negative. 
Furthermore consider a particle charged both with respect to the scalar and 
electromagnetic field. Then, by virtue of electric charge conservation, this particle can not just 
disappear when its effective mass becomes zero. 

One could ask as well the same question of classical radiation
in AdS space. It seems that a free floating particle
in AdS space will emit radiation as well. First, let
us stress that AdS space, unlike de Sitter, has a
globally defined time--like Killing vector.
Hence, there is energy conservation in this space and one can
diagonalize the free Hamiltonian for any field theory in AdS space for all
times. In the corresponding basis the analog of the amplitude
\eqref{ampmasshel} for AdS space
will vanish. Second, the answer to this question depends on the boundary
conditions at spatial infinity of AdS space, because of the
impossibility to define a Cauchy surface in such a background
\cite{Hawking:1973uf}. 

The radiation amplitude for de Sitter space given in equation
\eqref{ampmasshel} is small for realistic parameters, unless, as
during the early stage of the Universe, the vacuum energy is huge.
Thus the probability to create massive particles is small, but if
de Sitter space is eternal, then one can build a
perpetuum mobile and by living in such a space long enough extract
any amount of energy from the radiating charges. For example, one can 
construct
a box with walls which act as mirrors for the fermion and boson fields. Then 
one can put in such a
box some amount of massive fermions. Once in a while these fermions will
produce a massive scalar. By waiting long enough one can heat the box to any
extent. 

One might be tempted to think that if the back reaction on the radiation
were taken into account then one would see that the system of non--linear equations \eqref{system} has
some basis of stable soliton--like solutions for the fermions and bosons. 
Then fermions would not excite bosons. This is just equivalent to the statement 
that, while one can not diagonalize the free classical
Hamiltonian, one can somehow find a soliton--like basis of solutions of the
full interacting theory, in terms of which the full interacting Hamiltonian 
will be diagonalizable for all times. However, this statement, if it is true,
is a non--trivial assertion which would have to be shown explicitly. 

One could say that our conclusions are not surprising, since there is no
energy conservation in de Sitter space, because of the absence of a globally 
defined
time--like Killing vector and/or the absence of the corresponding
Casimir operator of the isometry group, which is responsible for
the time translations in de Sitter space. But then one has to
decide what is correct: either de Sitter space is not eternal and
the vacuum energy will eventually go to zero or one can heat ones
house for free. 

\section{Acknowledgments} 

AET would like to thank V.Zakharov, A.Rosly and O.Kancheli for
valuable discussions. Especially AET would like to thank A.Roura
for insisting on the further clarifications. The work was partially
supported by the
Federal Agency of Atomic Energy of Russian Federation and by the
grant for scientific schools NSh-679.2008.2. DS is supported by a
2008-2009 Fulbright Scholars Grants.


\begin{thebibliography}{99} 

\bibitem{BornerDurr}
G.Borner and H.P.Durr, Il Nuovo Cimento, Vol. LXIV A, No. 3 (1969). 

\bibitem{Kim:2002uz}
Y.~B.~Kim, C.~Y.~Oh and N.~Park, arXiv:hep-th/0212326. 

\bibitem{Lifshitz:1945du}
E.~Lifshitz, J.\ Phys.\ (USSR) {\bf 10}, 116 (1946). 

\bibitem{Myrhvold}
N.Myrhvold, {\it Phys. Rev.}, {\bf D 28} (1983) 2439. 

\bibitem{Akhmedov:2008pu}
E.~T.~Akhmedov and P.~V.~Buividovich,
Phys.\ Rev.\  D {\bf 78}, 104005 (2008)
[arXiv:0808.4106 [hep-th]]. 

\bibitem{Tsamis:1996qq}
N.~C.~Tsamis and R.~P.~Woodard,
Commun.\ Math.\ Phys.\  {\bf 162}, 217 (1994).
N.~C.~Tsamis and R.~P.~Woodard, Phys.\ Lett.\  B {\bf 292}, 269 (1992).
N.~C.~Tsamis and R.~P.~Woodard,
Class.\ Quant.\ Grav.\  {\bf 11}, 2969 (1994). N.~C.~Tsamis and
R.~P.~Woodard, Annals Phys.\  {\bf 238}, 1 (1995). 

\bibitem{Bros:2009bz}
J.~Bros, H.~Epstein, M.~Gaudin, U.~Moschella and V.~Pasquier,
arXiv:0901.4223 [hep-th]. J.~Bros, H.~Epstein and U.~Moschella,
arXiv:0812.3513 [hep-th].
J.~Bros, H.~Epstein and U.~Moschella,
JCAP {\bf 0802}, 003 (2008)
[arXiv:hep-th/0612184]. 

\bibitem{Tsaregorodtsev:1998mq}
L.~I.~Tsaregorodtsev and N.~N.~Medvedev,
Grav.\ Cosmol.\  {\bf 4}, 234 (1998)
[arXiv:gr-qc/9811072]. 


\bibitem{Poisson}
  L.~M.~Burko, A.~I.~Harte and E.~Poisson,
  Phys.\ Rev.\  D {\bf 65}, 124006 (2002)
  [arXiv:gr-qc/0201020]. 


\bibitem{Bousso:2001mw}
R.~Bousso, A.~Maloney and A.~Strominger, Phys.\ Rev.\  D {\bf 65},
104039 (2002) [arXiv:hep-th/0112218]. 

\bibitem{Mottola:1984ar}
E.~Mottola, Phys.\ Rev.\  D {\bf 31}, 754 (1985). 

\bibitem{Sadofyev} E.T.Akhmedov, A.Roura and A.Sadofyev, ``Classical
radiation of massless fields by free
falling charges in de Sitter space", to appear. 

\bibitem{Hawking:1973uf}
S.~W.~Hawking and G.~F.~R.~Ellis,
{\it  Cambridge University Press, Cambridge, 1973} 

\end{thebibliography}
\end{document}